\documentclass[preprint]{sigplanconf}

\usepackage{amsmath}

\usepackage{pstricks}
\usepackage{xspace}
\usepackage{graphicx}

\usepackage{bold-extra}
\usepackage{listings}
\usepackage{fancybox}

\usepackage{url}


\lstdefinelanguage{JavaScript}{
  keywords={break, case, catch, continue, debugger, default, delete, do, else, finally, for, function, if, in, instanceof, new, return, switch, this, throw, try, typeof, var, void, while, with},
  morecomment=[l]{//},
  morecomment=[s]{/*}{*/},
  morestring=[b]',
  morestring=[b]",
  sensitive=true
}

\newcommand{\TODO}[1]{}
\renewcommand{\TODO}[1]{{\bfseries\color{red} TODO: {#1}}}

\begin{document}

\setlength{\pdfpageheight}{\paperheight}
\setlength{\pdfpagewidth}{\paperwidth}

\conferenceinfo{CONF 'yy}{Month d--d, 20yy, City, ST, Country}
\copyrightyear{20yy}
\copyrightdata{978-1-nnnn-nnnn-n/yy/mm} 
\doi{nnnnnnn.nnnnnnn}


\title{Extending Basic Block Versioning with Typed Object Shapes}

\authorinfo{Maxime Chevalier-Boisvert}
           {DIRO, Universit\'e de Montr\'eal, Quebec, Canada}
           {chevalma@iro.umontreal.ca}
\authorinfo{Marc Feeley}
           {DIRO, Universit\'e de Montr\'eal, Quebec, Canada}
           {feeley@iro.umontreal.ca}

\maketitle



\begin{abstract}
Typical JavaScript (JS) programs feature a large number of object property
accesses. Hence, fast property reads and writes are crucial for good
performance. Unfortunately, many (often redundant) dynamic checks are implied
in each property access and the semantic complexity of JS makes it difficult
to optimize away these tests through program analysis.

We introduce two techniques to effectively eliminate a large proportion of
dynamic checks related to object property accesses. {\em Typed shapes} enable
code specialization based on object property types without potentially
complex and expensive analyses. {\em Shape propagation} allows the elimination
of redundant shape checks in inline caches. These two techniques combine
particularly well with Basic Block Versioning (BBV), but should be easily
adaptable to tracing Just-In-Time (JIT) compilers and method JITs with type
feedback.

To assess the effectiveness of the techniques presented, we have implemented
them in Higgs, a type-specializing JIT compiler for JS. The techniques are
compared to a baseline using Polymorphic Inline Caches (PICs), as well as
commercial JS implementations. Empirical results show that across the
26\unskip benchmarks tested, these
techniques eliminate on average 48\unskip\% of
type tests, reduce code size by 17\unskip\% and
reduce execution time by 25\unskip\%. On several
benchmarks, Higgs performs better than current production JS virtual machines.

\end{abstract}

\section{Introduction}\label{sec:intro}



Typical JavaScript programs make heavy use of object property accesses.
Unfortunately, the highly dynamic semantics of JS make optimization difficult
(Section~\ref{sec:js_objects}). Late binding, dynamic code loading and the
eval construct make type analysis a hard problem. Having to merge values of
multiple different types causes a loss of precision which is difficult to
avoid, even in analyses with high context-sensitivity.

Basic Block Versioning (BBV)~\cite{bbv_ecoop} is a Just-In-Time (JIT)
compilation strategy which allows rapid and effective generation of
type-specialized machine code without a separate type analysis pass or
complex speculative optimization and deoptimization
strategies~(Section~\ref{sec:bbv}). However, BBV, as previously
introduced, is inefficient in its handling of object property types.

The first contribution of this paper is the extension of BBV
with {\em typed object shapes} (Section~\ref{sec:typed_shapes}),
object descriptors which encode type information about object properties.
Type meta-information associated with object properties then becomes available
at property reads. This allows eliminating run-time type tests dependent
on object property accesses. The target of method calls is also known in most
cases.

The second contribution of this paper is a further extension of BBV with
{\em shape propagation} (Section~\ref{sec:shape_prop}), the propagation and
specialization of code based on object shapes. This allows eliminating some
redundant shape tests, making for more compact and more efficient machine
code. Shape propagation also provides some basic aliasing information.

Typed shapes and shape propagation were implemented in Higgs, a JIT compiler
for JS (ECMAScript 5.1) built around BBV. A detailed evaluation
of the performance implications
is provided in Section~\ref{sec:evaluation}. Empirical results across 26
benchmarks show that, on average, the techniques introduced eliminate
48\unskip\% of type tests, reduce code size by
17\unskip\% and reduce execution time by
25\unskip\%.

\section{Background}

\subsection{JavaScript Objects\label{sec:js_objects}}

JavaScript objects use a prototype-based inheritance model~\cite{es5_spec}
inspired from Self~\cite{self}. Objects can dynamically grow, meaning that
properties can be added to or deleted from an object at any time. The types of
properties are not constrained, and properties can be redefined to have any
type at any time. Semantically, JS objects can be thought of as behaving
somewhat like hash tables, but the semantics of property accesses are complex.

In JS, object properties can be plain values or accessor
(getter/setter) methods which may produce
side-effects when executed. Individual object properties can have read-only
(constant) attribute flags set, which prevents their redefinition.
When a property is not defined on an object, the lookup must traverse the
prototype chain recursively. These factors mean that each JS
property read or write implies multiple hidden dynamic tests. Ideally, most of
these tests should be optimized away to maximize performance.

Global variables in JS are stored on a first-class global object, which behaves
like any other. Properties of the global object can thus also be defined to
be read-only, or be accessor methods. Hence, optimizing global property
accesses is also a complex problem. Since the global object is a singleton and
typically large in size, modern JS engines such as Google's V8 tend to implement
it using a different strategy from regular objects.

\subsection{Object Shapes}

JS objects can be thought of as behaving like hash maps associating property
name strings to property values and attribute flags. However, implementing
objects using hash maps is inefficient both in terms of memory usage and
property access time. Doing so means that each object must store a name
string, a value and attribute flags for each property. Furthermore, each
property access must execute a costly hash table lookup which may involve
repeated indirections.

High-performance JS engines (V8, SpiderMonkey, etc.) rely on the concept of
object shapes, also known as ``hidden classes''. This approach aims to exploit
the fact that programs typically create many objects with the same properties,
that objects are usually initialized early in their lifetime, and that
property deletions and additions after initialization are infrequent.

\begin{figure}[tb]
\begin{center}
\includegraphics[scale=0.45]{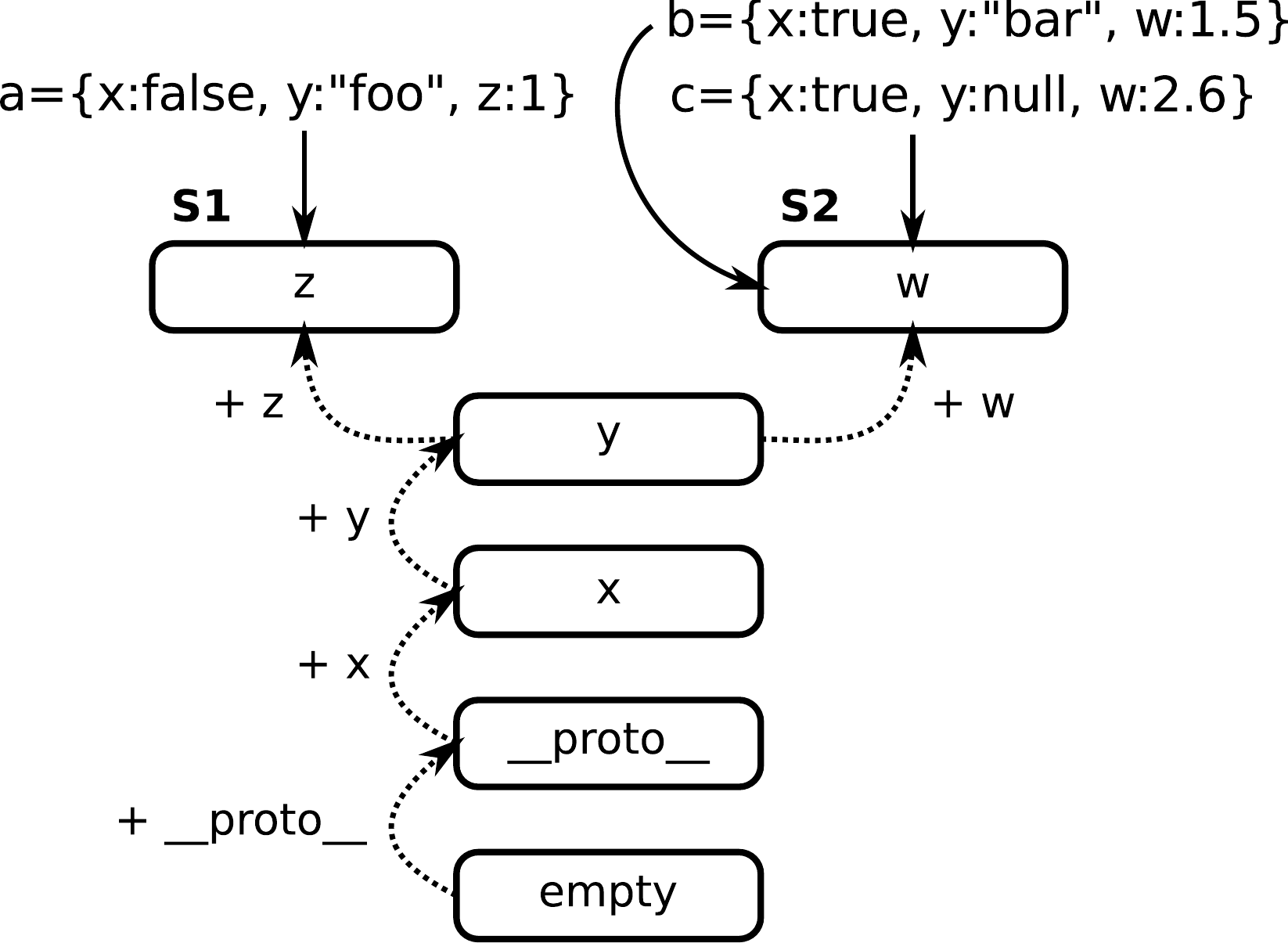}
\end{center}
\caption{Object shapes as part of a shape tree\label{fig:shape-tree}}
\end{figure}

Shapes are object layout descriptors. They are composed of
shape nodes, with each shape node containing the name, memory offset and
attribute flags for one property. All existing shape nodes are part of a global
tree structure representing the order in which properties have been added to
objects. Each object has a shape pointer which points to a shape node
representing the last property added to the said object.
All objects are initially created with no properties and begin their lifetime
with the empty shape. Adding a property to an object updates its shape pointer.

Figure~\ref{fig:shape-tree} illustrates the shape nodes for three different
JS object literals. All objects have a hidden {\tt \_\_proto\_\_} property
which stores a pointer to the prototype object. All three objects also share
properties named {\tt x} and {\tt y}, hence, part of the shapes of these two
objects are made of the same shape nodes. The last property added to object
{\tt a} is {\tt z}, and so it has shape {\tt S1}. The last property added to
objects {\tt b} and {\tt c} is {\tt w}, and so these have shape {\tt S2}.

\subsection{Polymorphic Inline Caches}\label{sec:pic}

Object shapes solve the space efficiency problem, that is, they are more space
efficient than using hash maps, since multiple objects with the same set of
properties and initialization order can share the same shape. However, shapes,
by themselves, do not make property accesses faster. Naively traversing the
shape structure of an object on every property access is likely slower
than implementing objects using hash maps.

\begin{figure}[tb]
\begin{center}
\includegraphics[scale=0.45]{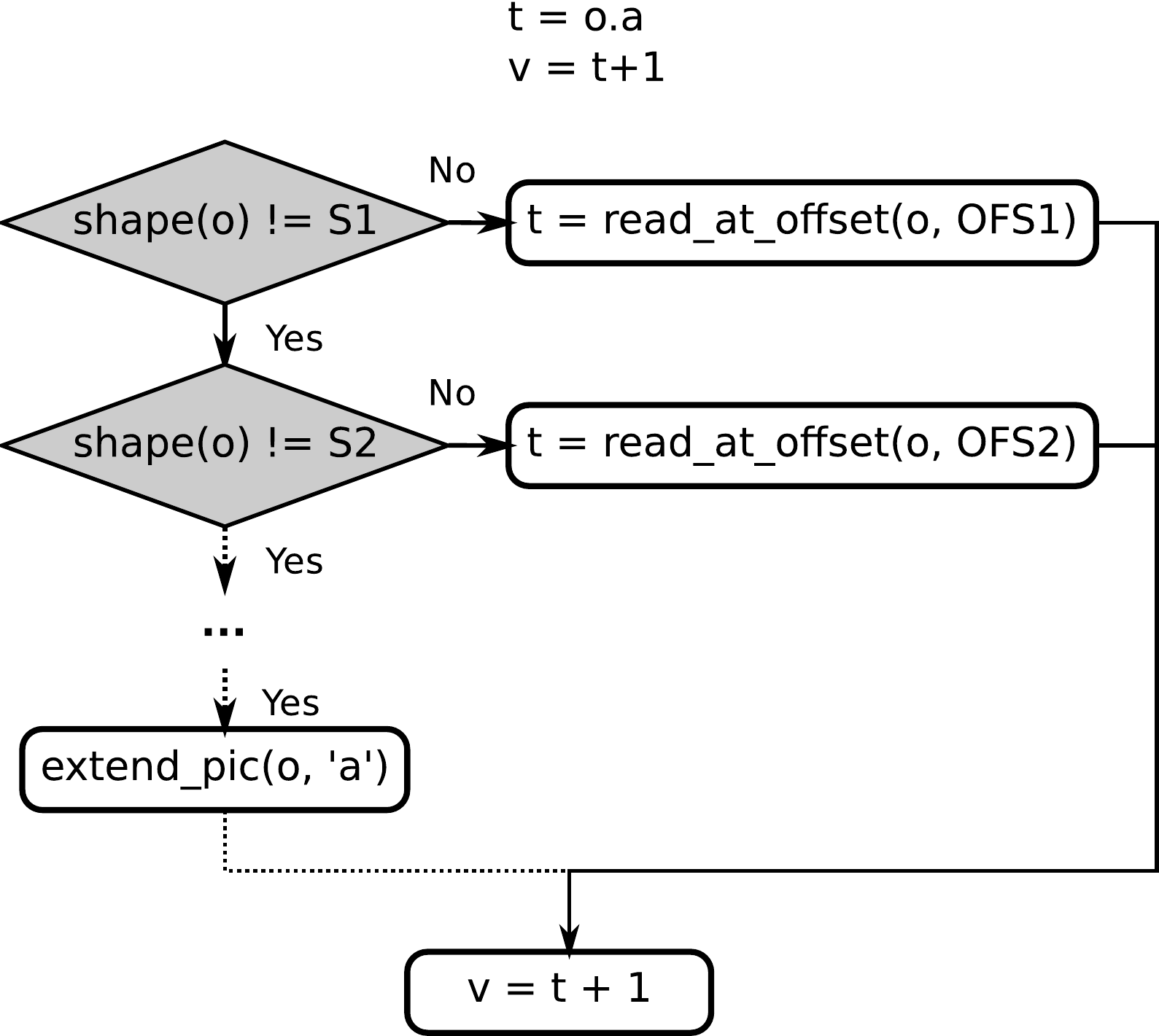}
\end{center}
\caption{Property read using a Polymorphic Inline Cache (PIC)\label{fig:pic}}
\end{figure}

Polymorphic Inline Caches (PICs), pioneered in the implementation of the Self
programming language~\cite{self, self_pic}, are commonly used to accelerate
property accesses. They are used by modern JS VMs
such as V8 and SpiderMonkey. The core idea is to generate machine code
on-the-fly to determine the shape of an object and generate an efficient
dispatch at each property access site. This machine code takes the form of a
cascade of shape test operations and is updated as new object shapes are
encountered. PICs can be thought of as offloading the property lookup overhead
to code generation time instead of execution time.

Figure~\ref{fig:pic} illustrates a property read implemented using a PIC.
Two shape tests match previously encountered object shapes. Each test, if it
encounters a matching shape, triggers the execution of a load machine
instruction which reads the property from the object at the correct memory
offset. This memory offset is determined at code generation time based on the
object's shape, which tells us where each property is located. In the optimal
case for a PIC, a property read can be as fast as one comparison and one load
machine instruction.

\subsection{Basic Block Versioning\label{sec:bbv}}

Basic Block Versioning (BBV), as introduced by Chevalier-Boisvert and
Feeley~\cite{bbv_ecoop} and adapted to Scheme by Saleil and
Feeley~\cite{vers_scheme} is a simple JIT compilation technique resembling
trace compilation. BBV generates efficient type-specialized code without the
use of costly type inference analyses or profiling. Basic blocks are
lazily cloned and specialized on-the-fly in a way that allows the compiler
to accumulate type information while machine code is generated, without a
separate type analysis pass. The accumulated information allows the removal
of redundant type tests, particularly in performance-critical paths.

BBV lets the execution of the program itself drive the
generation of type-specialized code, and is able to avoid some of the
precision limitations of traditional, conservative type analyses as well as
avoiding the implementation complexity of speculative optimization techniques.
BBV does not require the use of on-stack replacement or deoptimization. It is
intended to generate type-specialized code at low overhead, without
needing a fixed point type analysis pass, which makes it particularly
attractive for baseline compilers.

Higgs segregates values into a few categories based on
type tags~\cite{type_tags}. These categories are: 32-bit integers
({\tt int32})\footnote{Note that while according to the ES5 specification
all JavaScript numbers are
IEEE double-precision floating point values, high-performance JavaScript
VMs typically attempt to represent small integer values
using machine integers so as to improve performance by using
lower latency integer arithmetic instructions. We have made the same design
choice for Higgs.}, 64-bit floating point values ({\tt float64}),
miscellaneous JS
constants ({\tt const}), and four kinds of garbage-collected pointers inside
the heap ({\tt string}, {\tt object}, {\tt array}, {\tt closure}). These
type tags form a simple, first-degree notion of types that is used to drive
code versioning.

BBV, as introduced, deals only with function parameter and local variable
types. It has no mechanism for handling object property types and global
variable types. The current work extends BBV to include a more advanced notion
of object types based on {\em typed shapes}, and enable type-specialization 
based on object property and global variable types.

\section{Typed Shapes}
In this section we present the main contributions of this paper.

\subsection{Typed Shapes and Property Types\label{sec:typed_shapes}}

\begin{figure*}[tb]
\begin{center}
\includegraphics[scale=0.45]{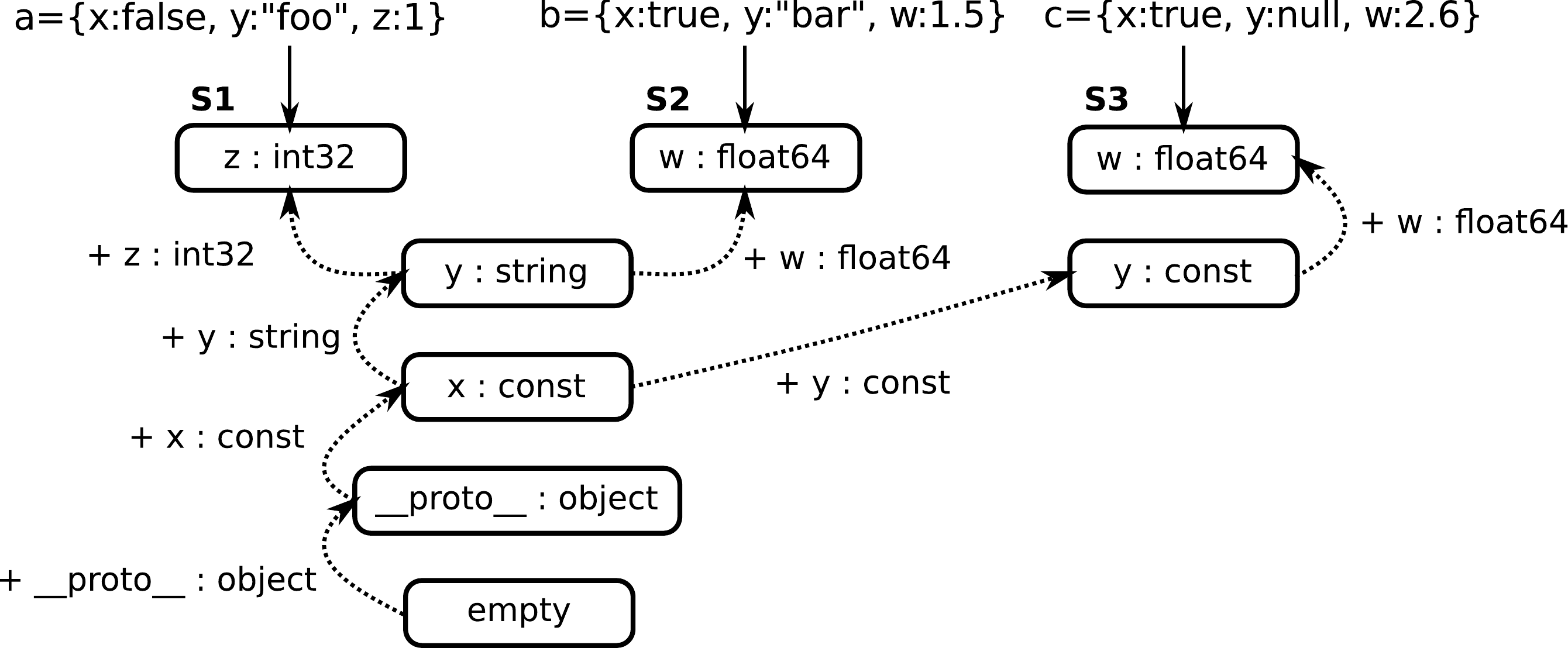}
\end{center}
\caption{Type meta-information on object shapes and property additions\label{fig:typed-shapes}}
\end{figure*}

Object shapes in other JS engines encode property names, slot indices and
meta-information such as attribute flags (writable, enumerable, etc.). We
extend shapes to also encode property types: this makes
it possible for us to specialize code based on the types of property values.
Testing the shape of an object once gives us the type of all its properties.

Figure~\ref{fig:typed-shapes} shows the object shapes associated with three
different JS object literals. Shape nodes are now annotated with type tags
corresponding to property values. Objects {\tt b} and {\tt c} share the same
property names, but the type of their {\tt y} property differs. The property
{\tt b.y} is a string, whereas {\tt c.y} has value {\tt null} which has type
tag {\tt const}.

Our definition of typed shapes is not recursive. Shapes corresponding to
property values which are object references do not encode the shape of the
object being referenced. This is because objects are mutable. Hence, if
{\tt a.b} is an object, its shape cannot be guaranteed to remain the same
during the execution of a program, but objects will always remain objects,
so the type tag of {\tt a.b} will not change so long as this property is
not overwritten.

With typed shapes, property values can always be stored in an unboxed
representation, thereby avoiding boxing and unboxing overhead.
In the optimal case, properties can be read and written in a single machine
instruction. In commercial JS engines such as V8, Floating Point (FP)
values may be stored in boxed representations, but Higgs can store FP
values inside objects without indirection.

A further advantage is that the shape of an object can tell us whether or not
the object has a prototype or not. This eliminates the need to perform a
{\tt null} check when going up the prototype chain during a property read.

\subsection{Method Identity and the Global Object}\label{sec:method}

\begin{figure}[tb]
\begin{center}
\includegraphics[scale=0.45]{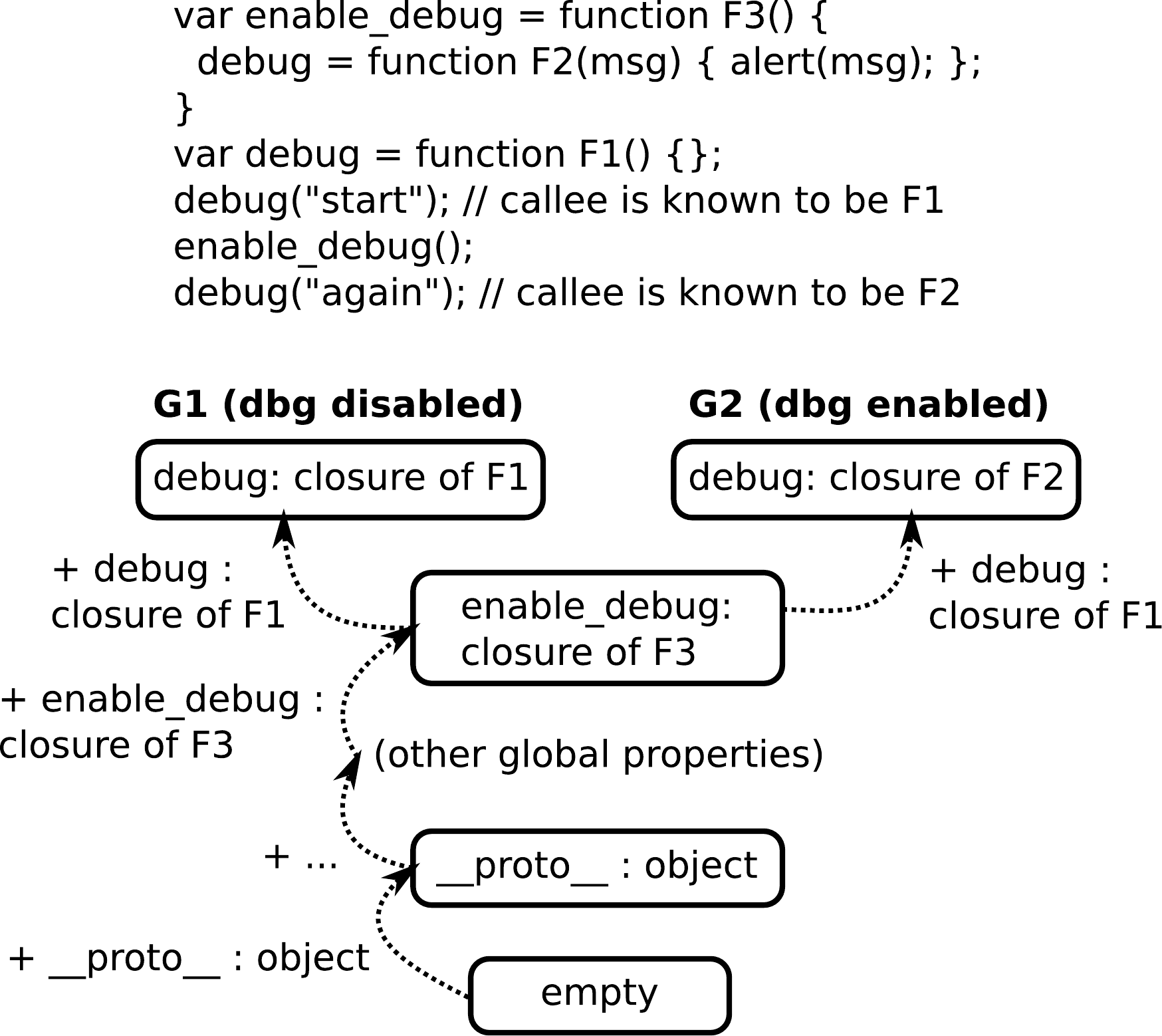}
\end{center}
\caption{Object shapes for global properties and functions\label{fig:global-shapes}}
\end{figure}

The property type information currently encoded in Higgs includes type tags,
but also function pointers (function/method identity). Encoding function
pointers makes it possible to know the identity of callees at call sites.
This enables us to specialize call sites based on the callee.
For instance, when the identity of a callee is known, passing
unused argument values (such as the hidden {\tt this} argument) can be
avoided.

Our approach uses a unified implementation for all objects, including the
global object. Hence, in Higgs, global property accesses can be optimized
using the same techniques as regular property accesses. This contrasts with
V8, which uses a collection of individual mutable cells to implement its
global object.

Figure~\ref{fig:global-shapes} illustrates the global object shape in relation
to a snippet of code where a call to {\tt enable\_debug} replaces an
inactive implementation of the {\tt debug} function by one which
displays error messages. The {\tt enable\_debug} function causes the global
object to switch from shape {\tt G1} to {\tt G2}. The global object shape
encodes the identity of functions, meaning that for both calls to
{\tt debug}, we know that it must be a function and what its identity is.

\subsection{Shape Propagation}\label{sec:shape_prop}

As described in Section~\ref{sec:pic}, Polymorphic Inline Caches (PICs) are
a lazily generated chain of dynamic tests to identify an object's
shape and quickly select a fast implementation of a property read or write.
We extend upon this idea, combining it with BBV,
so that code may be specialized based on the shape of an object. Shape tests
which are normally part of PICs are used to identify and propagate the
shape of objects. Propagating shapes allows eliminating redundant (repeated)
shape tests, and other optimizations based on an object's shape. For instance,
if we know that two object references point to objects of different shapes,
then we know that they cannot point to the same object.

As shown in~\cite{bbv_ecoop}, most Static Single Assignment (SSA) values are
monomorphic in terms of type tags. Few values are polymorphic. This remains
true when shapes come into the picture. Most program points see only one shape
for a given value. However, the objects which are polymorphic in shape are
sometimes megamorphic. That is, one property access site can receive objects
of a large number of different shapes. This can quickly lead to combinatorial
explosions in the number of possible block versions.

Versioning serves to propagate type information effectively. Code duplication
is useful, so long as there is not too much of it. The cost
of tracking all possible types of megamorphic SSA values is not worthwhile,
since these values are fairly rare. Hence, we have taken the approach of
limiting how many different shapes can be tracked for a given SSA value.
The {\tt maxshapes} parameter serves to prevent code size explosions,
avoiding the situation where rare polymorphic values cause disproportional
code size growth.

\subsection{Shape Flips}

Overwriting an existing property value may cause an object to transition
to a new shape if the type of the new value doesn't match the type encoded in
the object's current shape. We call this a shape flip. For instance, if
object {\tt c} from Figure~\ref{fig:typed-shapes} was to have its {\tt y}
property overwritten with a string value (e.g. {\tt c.y = "bif"}), then the
shape of {\tt c}, which was previously {\tt S3}, would change to {\tt S2}.

Empirically, such shape flips are relatively rare (see
Section~\ref{sec:reads-writes}). However, property writes
still need to be guarded based on the type of the written value. Fortunately,
most of these guards are redundant and can be safely eliminated because BBV
is very effective at propagating value types, and so most value types are
known when machine code is generated.

Since most value types are known, and property reads tend to
significantly outnumber property writes (see Section~\ref{sec:reads-writes}),
the overhead of testing some value types before writing them is easily recouped
by eliminating type tests after property reads. That is, the number of type
tests eliminated after property reads vastly outnumbers the number of guards
added before property writes.

Changing an object's shape can be done with little execution time overhead.
Assuming that the new shape has been previously allocated and initialized,
which is often the case, changing an object's shape is only a matter of
writing a new shape pointer on the object, which can be done in a single
machine instruction.

\subsection{Shape and Type Checks}

\begin{figure}[tb]
\begin{center}
\includegraphics[scale=0.45]{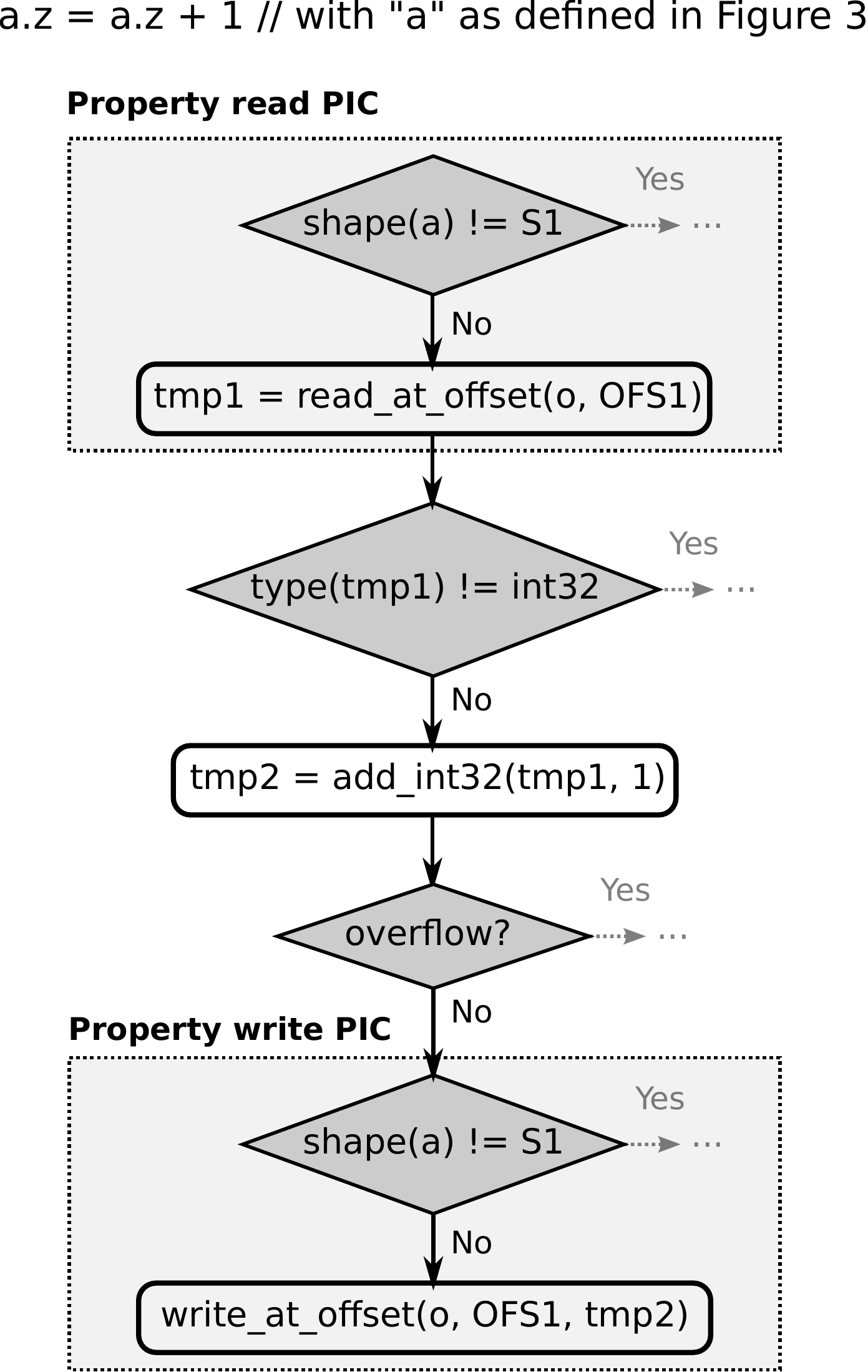}
\end{center}
\caption{Operations involved in a property read and write with traditional Polymorphic Inline Caches (PICs)\label{fig:ops-incr-pic}}
\end{figure}

\begin{figure}[tb]
\begin{center}
\includegraphics[scale=0.45]{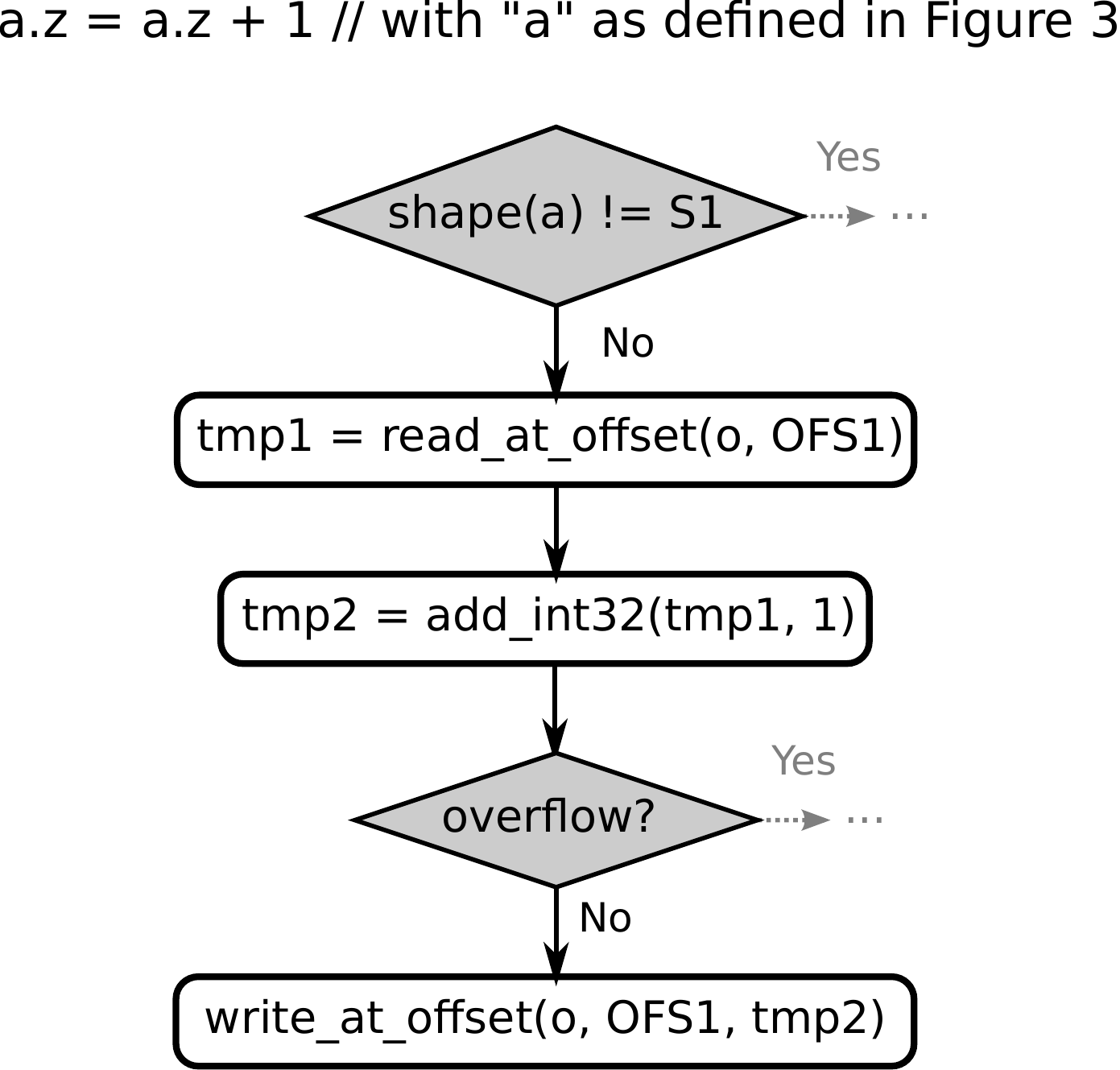}
\end{center}
\caption{Operations involved in a property read and write with typed shapes and shape propagation, starting with an object of unknown shape\label{fig:ops-incr-typed}}
\end{figure}

With polymorphic inline caches, reading or writing to an object property
implies first performing a number of dynamic checks to dispatch read or
write operations based on the object shape (see section~\ref{sec:pic}). Many
JS primitives, including arithmetic operators, also perform dynamic dispatch
based on value types.

Figure~\ref{fig:ops-incr-pic} illustrates the operations involved in
incrementing the value of an integer property on an object
({\tt a.z = a.z + 1}) when using traditional PICs (without typed shapes).
There are four dynamic checks. A first check is performed to dispatch based
on the object shape when reading the property. A second check is performed to
dispatch based on the type of the property's value (which is {\tt int32}
in this case). A third check is performed to verify that the result of the
integer addition operation did not result in an integer overflow. Finally, a
fourth dynamic check is performed when writing back the incremented value.
This last check is necessary because the property read and property write
PICs are distinct.

Typed shapes and shape propagation produce more efficient code, as
illustrated in Figure~\ref{fig:ops-incr-typed}. The dynamic
dispatch based on the property type is eliminated, because this type is
encoded in the object's shape, and is thus automatically known once the
object's shape has been tested. The dispatch based on the object's
shape when writing back the property is eliminated because the object's shape
was previously tested and this information is propagated to the write.
There is no need to guard the type of {\tt tmp2} when writing the new
property value because this type is deduced based on the type of {\tt tmp1}.

\section{Evaluation}\label{sec:evaluation}
\subsection{Experimental Setup}

We have tested an implementation of the Higgs JIT compiler implementing typed
shapes and shape propagation on a
total of 26\unskip classic benchmarks from the SunSpider and V8
suites. One benchmark from the SunSpider suite and one from the V8 suite were
not included in our tests because Higgs does not yet implement the required
features. Benchmarks making use of regular expressions were
discarded because unlike V8 and SpiderMonkey, Higgs does not implement JIT
compilation of regular expressions, and neither does
Truffle/JS~\cite{trufflejs, truffle}.

To measure execution time separately from compilation time in a manner
compatible with V8, SpiderMonkey, Truffle/JS and Higgs, we have modified
benchmarks so that they could be run in a loop. A number of warmup iterations
are first performed so as to trigger JIT compilation and optimization of code
before timing runs take place.

The number of warmup and timing iterations were scaled so that short-running
benchmarks would execute for at least 1000ms in total during both warmup and
timing. Unless otherwise specified, all benchmarks were run for at least 10
warmup iterations and 10 timing iterations.

V8 version 3.29.66, SpiderMonkey version C40.0a1, Truffle/JS v0.5 and GCC
version 4.7.3 were used for performance comparisons. Tests were executed on a
system equipped with an Intel Core i7-4771 quad-core CPU with 8MB L3 cache
and 16GB of RAM running Ubuntu Linux 12.04. Dynamic CPU frequency scaling was
disabled.

\subsection{More Shape Nodes}

\begin{figure}[tb]
\begin{center}
\begin{tabular}{l | c | c | c}
\hline
benchmark&untyped shapes&typed shapes&ratio\\
\hline
partial-sums&13&18&1.4\\
v8-raytrace&214&420&2.0\\
cordic&17&40&2.4\\
bitwise-and&2&5&2.5\\
3d-raytrace&65&164&2.5\\
bits-in-byte&2&6&3.0\\
nsieve-bits&3&9&3.0\\
3bits-byte&2&6&3.0\\
nsieve&3&9&3.0\\
deltablue&127&396&3.1\\
fannkuch&3&10&3.3\\
recursive&4&14&3.5\\
splay&56&200&3.6\\
nbody&35&128&3.7\\
spectral-norm&6&22&3.7\\
navier-stokes&45&186&4.1\\
fasta&28&123&4.4\\
3d-morph&7&34&4.9\\
binary-trees&16&100&6.2\\
richards&92&604&6.6\\
base64&8&56&7.0\\
crypto-sha1&23&200&8.7\\
crypto-md5&26&230&8.8\\
3d-cube&43&389&9.0\\
v8-crypto&289&3881&13.4\\
earley-boyer&535&32324&60.4\\
\hline
mean&64.0&1522.1&4.5\\
\hline
\end{tabular}

\end{center}
\caption{Shape nodes created with and without typed shapes\label{fig:num_shapes}}
\end{figure}

Typed shapes create shape nodes for each possible type a property may have.
This necessarily increases the number of shapes created over the course of
a program's execution (see Figure~\ref{fig:num_shapes}). Enabling typed shapes
results in a 4.5\unskipx mean increase in the number of shapes
created.

The case of {\tt earley-boyer} shows the most dramatic increase. This is due
to global object properties being redefined late during the benchmark's
execution, which causes large sections of the shape tree corresponding to
the global object to be regenerated. It may be possible to optimize shape tree
transformations and avoid recreating all shapes descending from the redefined
property shape, but as we will see in the next subsections, the increase in
the number of shape nodes created does not cause performance problems.

The peak memory usage of the Higgs process was measured with various
configurations. With typed shapes and unlimited shape propagation
({\tt maxshapes=$\infty$}), process memory usage increases by
3.7\unskip\% on average. The {\tt earley-boyer} benchmark,
despite the large increase in the number of shapes generated, has
lower memory usage than the baseline without typed shapes. This is because
shape nodes are relatively small and typed shapes allow a large reduction
in generated machine code size, as shown in Figure~\ref{fig:code_size}.

\subsection{Reads and Writes}\label{sec:reads-writes}


For most benchmarks, less than 4\% of total writes result in a shape flip,
and in the worst case, just 17\unskip\% of total
writes do. Writes to the global object are more likely to
result in a shape flip. This is because much of these writes are due to the
initialization of global variables which originally had the {\tt undefined}
value. Such initial writes to the global object constitute a minority of
total property writes.

The benchmarks in our set perform between 1.4\unskip and 
567\unskip property reads for every property write. Since
property reads outnumber property writes and shape flips have relatively little
overhead, our prediction was that the overhead of shape flips would be easily
recouped by the reduction in type test provided by typed shapes. This
prediction is confirmed by the results obtained. The {\tt richards} benchmark,
which has the highest relative occurrence of shape flips, is actually one of
the benchmarks which obtain the most significant speedup from typed shapes
(see Figure~\ref{fig:exec_time}).

\subsection{Dynamic Tests}

\begin{figure*}[tb]
    \begin{center}
    \includegraphics[scale=1.00]{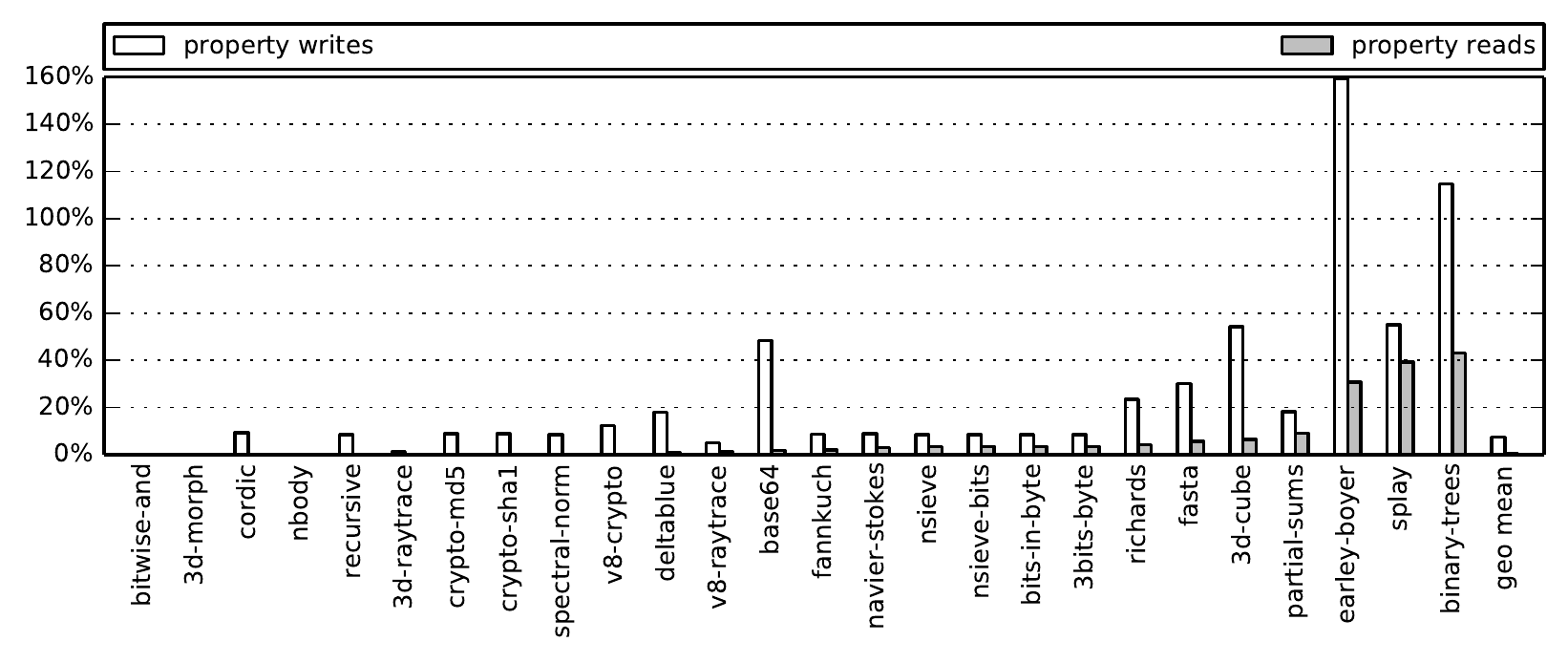}
    \end{center}
\caption{Number of type tag guards relative to property writes and reads\label{fig:num_tag_guards}}
\end{figure*}

Encoding type tags in property shapes implies that property writes must be
guarded with type tag checks. Figure~\ref{fig:num_tag_guards} shows the
number of executed property guards relative to the number of property writes
and property reads. In most cases, property
writes do not execute any guards. This is because with BBV, the type tag of
values is known and does not need to be tested.

There are rare cases, such as with {\tt earley-boyer}, where the
number of tag guards outnumber property writes. In this case, it is because
this benchmark is the output to a Scheme-to-JS compiler, and creates
highly polymorphic {\tt cons} pairs with elements of many different types
through a unique constructor function. Even in this case, however, the number
of write guards is less than the number of property reads, suggesting
that the additional cost paid when testing type tags before property writes
will be offset by the savings of eliminating type tests after property reads.

\begin{figure*}[tb]
    \begin{center}
    \includegraphics[scale=1.00]{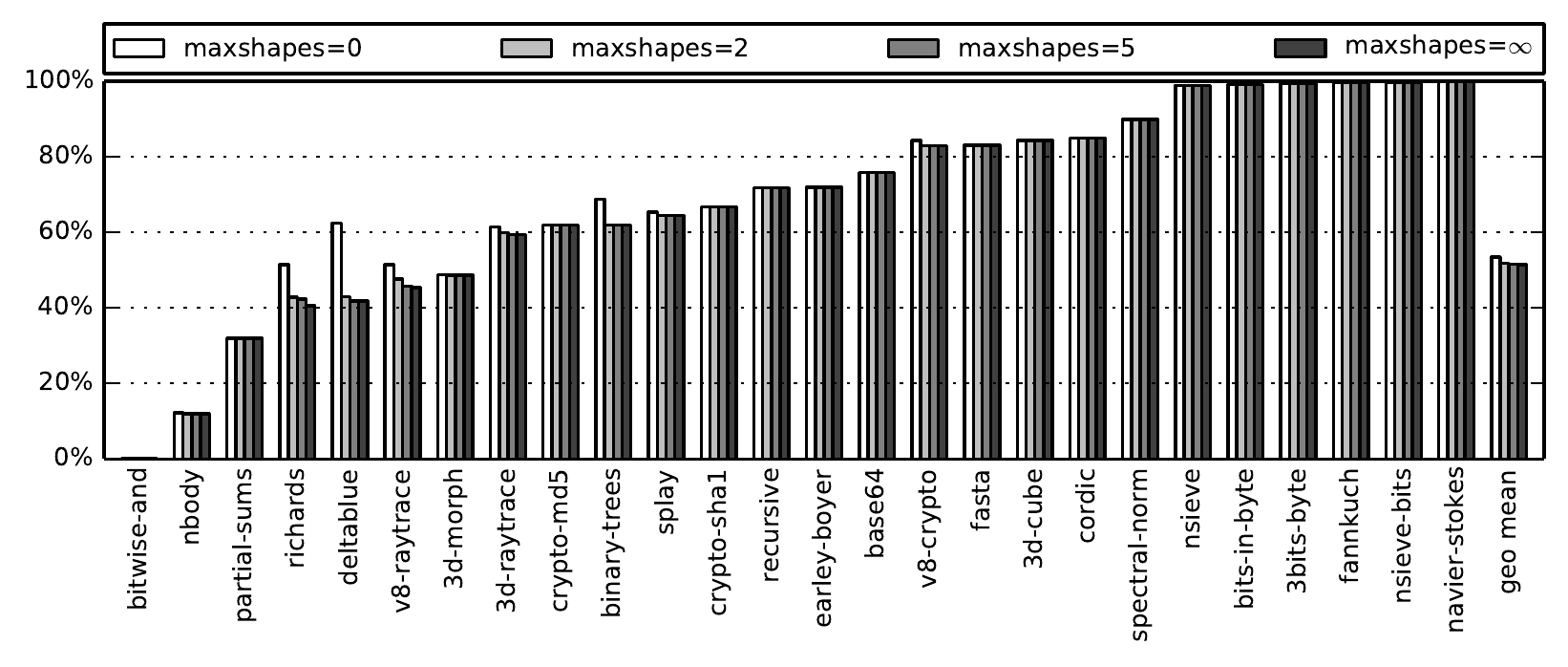}
    \end{center}
\caption{Number of type tests relative to inline cache baseline\label{fig:num_type_tests}}
\end{figure*}

Figure~\ref{fig:num_type_tests} shows the total number of type tag tests
(including guards on property writes) performed with different {\tt maxshapes}
parameter values relative to a baseline which uses traditional inline caches
without typed shapes or shape propagation. The chart makes it clear that typed
shapes can reduce the number of type tests very significantly. In the case of
the {\tt bitwise-and} microbenchmark, which operates entirely on two
global variables, type tests are reduced by nearly 100\%. On average, a
reduction of 48\unskip\% is obtained with
{\tt maxshapes=2}.

Note that going from {\tt maxshapes=0} to higher values produces a
slight reduction in type tests on some benchmarks. This is because enabling
the propagation of shapes allows eliminating a {\tt null} check while
traversing the prototype chain, since the prototype link is itself represented
as a typed property. The benchmarks which benefit the most from this phenomenon
are those which make heavy use of prototypal inheritance.

\begin{figure*}[tb]
    \begin{center}
    \includegraphics[scale=1.00]{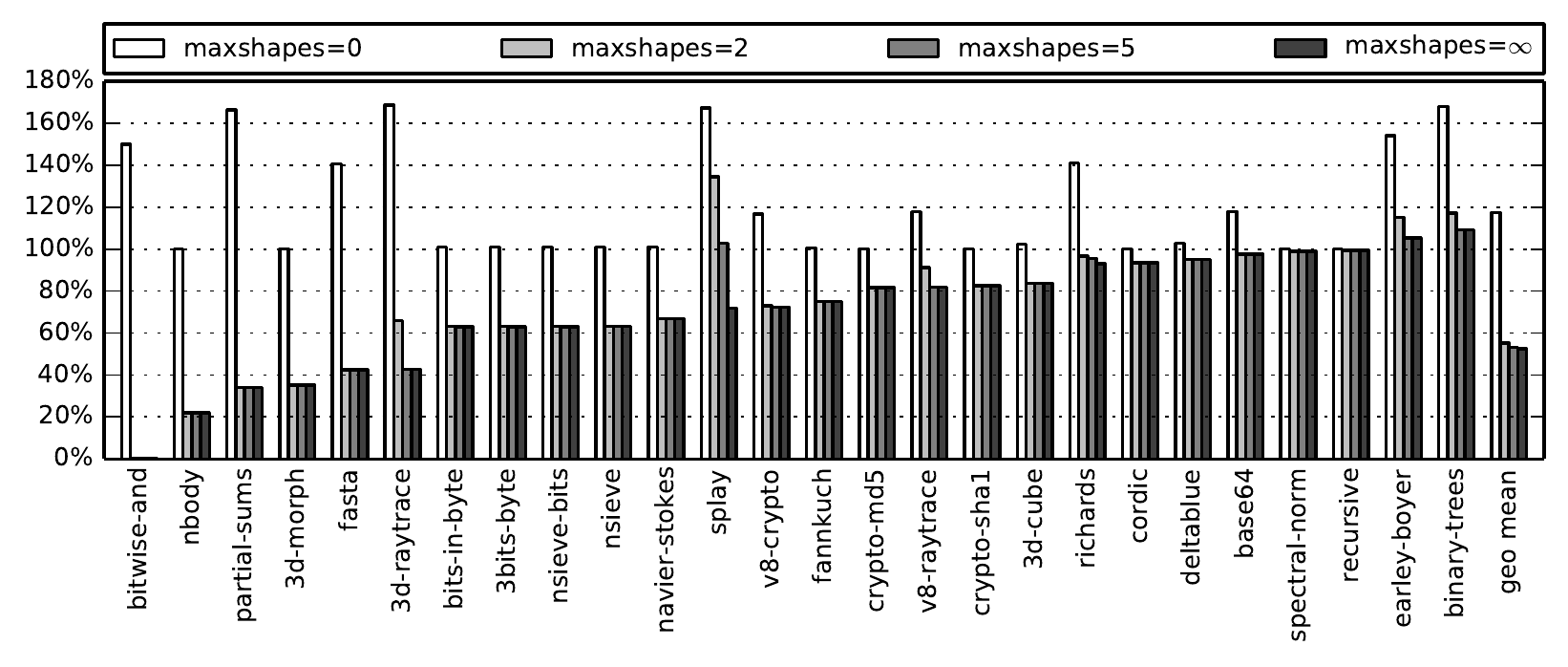}
    \end{center}
\caption{Number of shape tests relative to inline cache baseline\label{fig:num_shape_tests}}
\end{figure*}

Figure~\ref{fig:num_shape_tests} illustrates the number of shape tests relative
to a baseline using inline caches without typed shapes or shape propagation.
Notably here, setting {\tt maxshapes=0} increases the number of
shape tests in many cases, by 17\unskip\% on
average. This is because typed shapes result in the creation of more shape
nodes, as shown in Figure~\ref{fig:num_shapes}. Hence, individual inline
caches tend to produce longer chains of tests.

Interestingly, setting {\tt maxshapes=2} produces a reduction in the
number of shape tests in most cases
(45\unskip\% on average). This is because there
are many instances where multiple property reads on the same object occur
within a given function, and shape propagation can allow us to eliminate
further shape tests after the first property access on an object.

\subsection{Function Calls}


In the absence of typed shapes, Higgs does not know the identity of callees at
most call sites. With typed shapes, on average, callee identity is known for
90\unskip\% of calls. For most benchmarks, the
identity is known for all calls. There are some exceptions because at
present Higgs cannot specialize calls performed using {\tt apply} or calls
made using closures passed as function arguments.

\subsection{Code Size}

\begin{figure*}[tb]
    \begin{center}
    \includegraphics[scale=1.00]{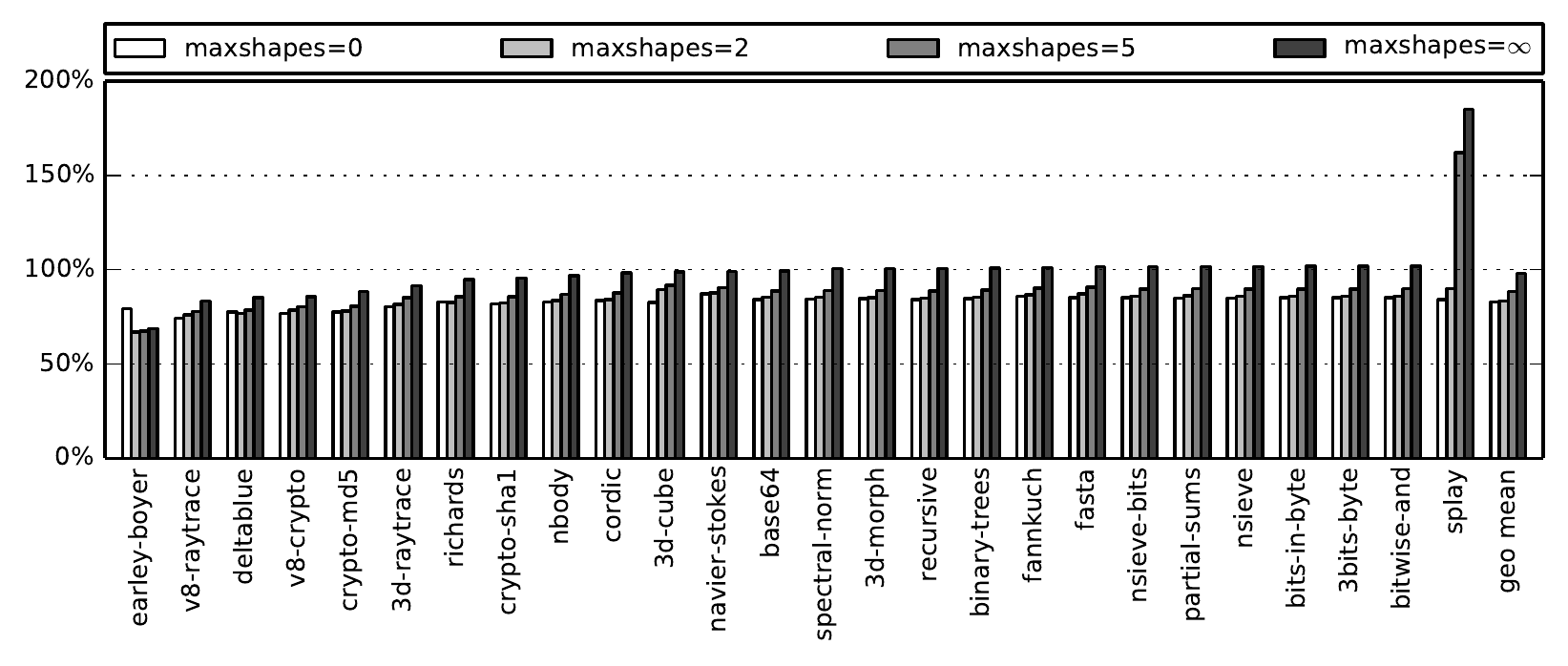}
    \end{center}
\caption{Code size relative to inline cache baseline\label{fig:code_size}}
\end{figure*}

Figure~\ref{fig:code_size} shows the effect of typed shapes with various
{\tt maxshapes} parameter values on machine code size. With
{\tt maxshapes=0}, typed shapes result in a smaller code size on every
benchmark, with an average reduction of
17\unskip\%. However, enabling shape propagation
without a limit on the number of shapes propagated ({\tt maxshapes=$\infty$})
results in a code size increase in several cases, and a pathological code
size blowup in the case of the {\tt splay} benchmark. Setting {\tt maxshapes=2}
yields a 17\unskip\% average code size reduction
and avoids the pathological code size blowup on the {\tt splay} benchmark.

\subsection{Compilation time}

\begin{figure*}[tb]
    \begin{center}
    \includegraphics[scale=1.00]{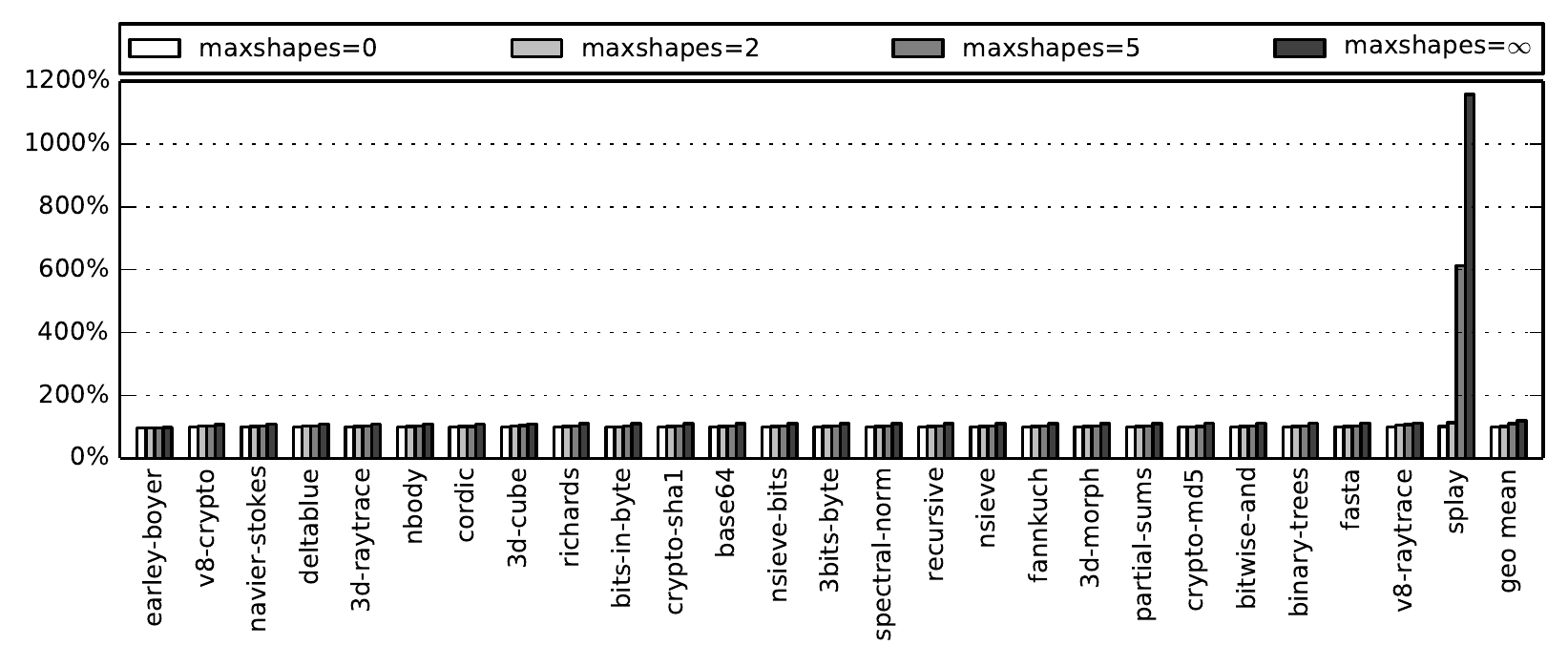}
    \end{center}
\caption{Compilation time relative to inline cache baseline\label{fig:comp_time}}
\end{figure*}

Typed shapes result in the allocation and manipulation of more shape nodes,
which can add compilation-time overhead. The techniques
presented in this paper may also result in the generation of more
machine code, which can also increase compilation times. 

The effect of typed shapes and shape propagation on compilation time are
explored in Figure~\ref{fig:comp_time}. The {\tt splay} benchmark, which
exhibits a pathological code size blowup when shape propagation is left
unlimited ({\tt maxshapes=$\infty$}), also shows a compilation time blowup.
However, once again, setting {\tt maxshapes=2} resolves the issue.
With {\tt maxshapes=2}, the mean compilation time increase is just
1\unskip\%, as upposed to
20\unskip\% with {\tt maxshapes=$\infty$}.

\subsection{Execution Time}

\begin{figure*}[tb]
    \begin{center}
    \includegraphics[scale=1.00]{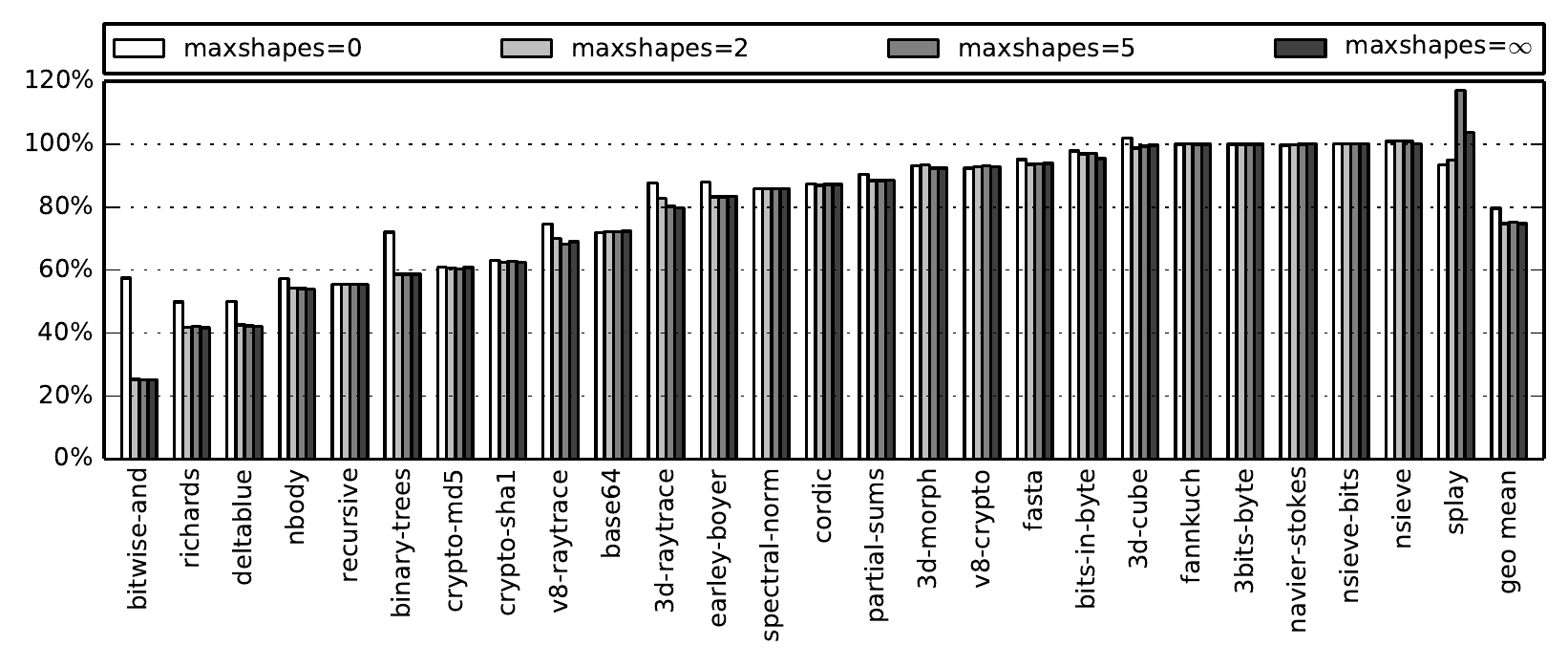}
    \end{center}
\caption{Execution time relative to inline cache baseline (lower is better)\label{fig:exec_time}}
\end{figure*}

Figure~\ref{fig:exec_time} shows the execution time with different
{\tt maxshapes} parameter values relative to a baseline using inline caches
only. Setting {\tt maxshapes=0} produces an average execution time reduction
of 20\unskip\%, compared to
25\unskip\% when {\tt maxshapes=2}. Setting
{\tt maxshapes} to higher values produces speedups on most
benchmarks, but results in a performance degradation on the {\tt splay}
benchmark.

The {\tt splay} benchmark has a high degree of shape polymorphism, and
illustrates the motivation for the {\tt maxshapes} parameter. Unlimited shape
propagation can result in code bloat in some cases (as illustrated in
Figure~\ref{fig:code_size}) which may increase instruction cache
misses and cause performance degradations.

The {\tt bitwise-and} microbenchmark, using only global variable accesses in
a small loop with no shape polymorphism, is an ideal showcase for typed shapes
and shape propagation. However, in the current implementation of Higgs, shape
propagation produces significant performance gains on a minority of sizable
benchmarks. We believe this is in large part because the current implementation
cannot preserve known shapes across function calls. Overcoming this limitation
is part of future work (see Section~\ref{sec:future}).

\subsection{Comparison with V8}

\begin{figure*}[tb]
    \begin{center}
    \includegraphics[scale=1.00]{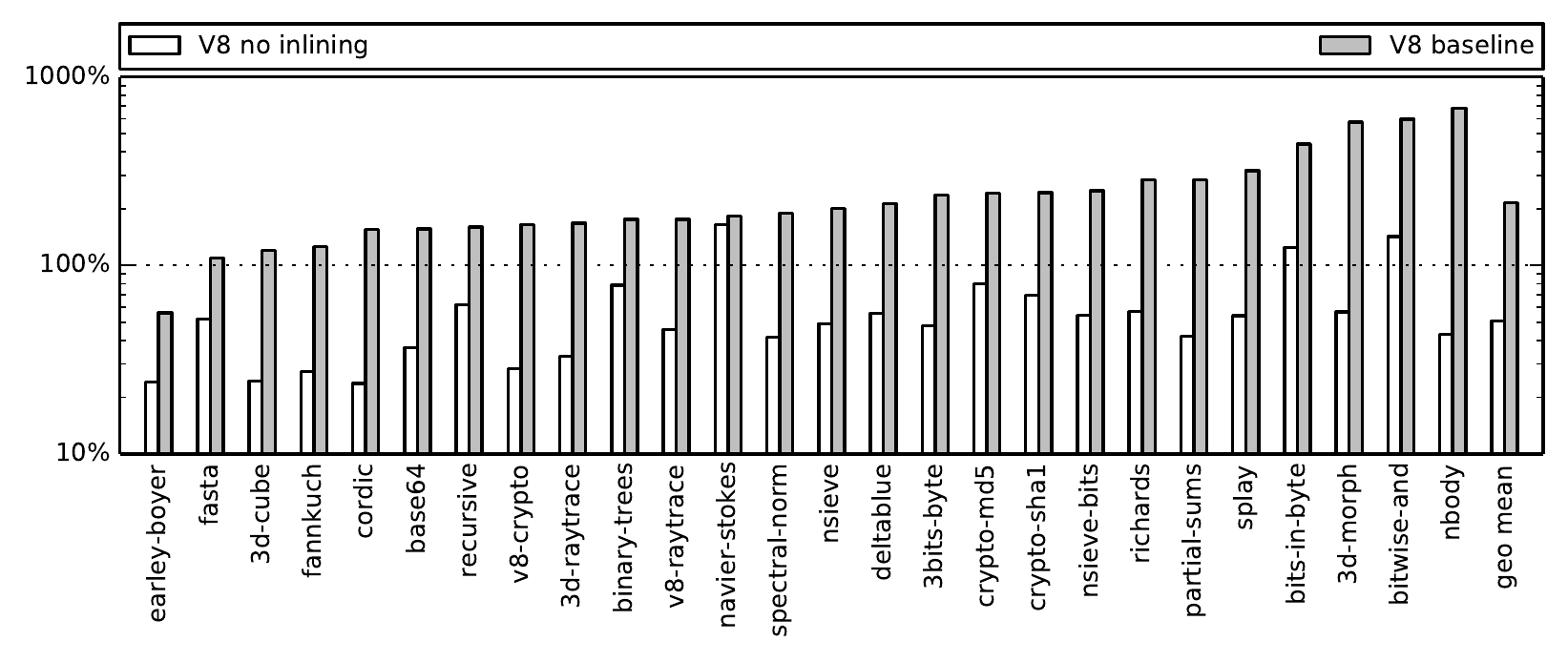}
    \end{center}
\caption{Execution time relative to V8 baseline and V8 without inlining (log scale, bars above 100\% favor Higgs)\label{fig:against_v8}}
\end{figure*}

The performance of Higgs ({\tt maxshapes=2}) was compared
with that of two configurations of Google's V8 JIT compiler (see
Figure~\ref{fig:against_v8}). The first configuration uses only the V8 baseline
JIT (with Crankshaft disabled). The second configuration uses Crankshaft,
but disables inlining to make its capabilities more comparable to that of Higgs.

With the addition of typed shapes and shape propagation, Higgs easily
outperforms the V8 baseline compiler on all but one out of 26\unskip
benchmarks, with speedups of up to 682\unskip\%.
This suggests that the quality of the code generated by Higgs is much superior
to the V8 baseline compiler. The large speedups obtained on {\tt bitwise-and}
show that Higgs has much faster global variable accesses than the V8 baseline
compiler.

Higgs lags behind Crankshaft without inlining, with Crankshaft
performing 49\unskip\% better on average. This
is not surprising since Crankshaft is able to use type feedback and
sophisticated analyses to optimize code at a much higher level than that of
Higgs. Higgs is also at a disadvantage because it does not perform efficient
register allocation for floating-point values, instead shuffling them in and
out of general-purpose registers. Furthermore, we surmise based on the
performance obtained on the {\tt recursive} microbenchmark that Crankshaft
has better optimized function calls than Higgs. This may have a significant
performance impact on several benchmarks. Optimizing the calling convention
used by Higgs is part of future work.

\subsection{Comparison with SpiderMonkey}

\begin{figure*}[tb]
    \begin{center}
    \includegraphics[scale=1.00]{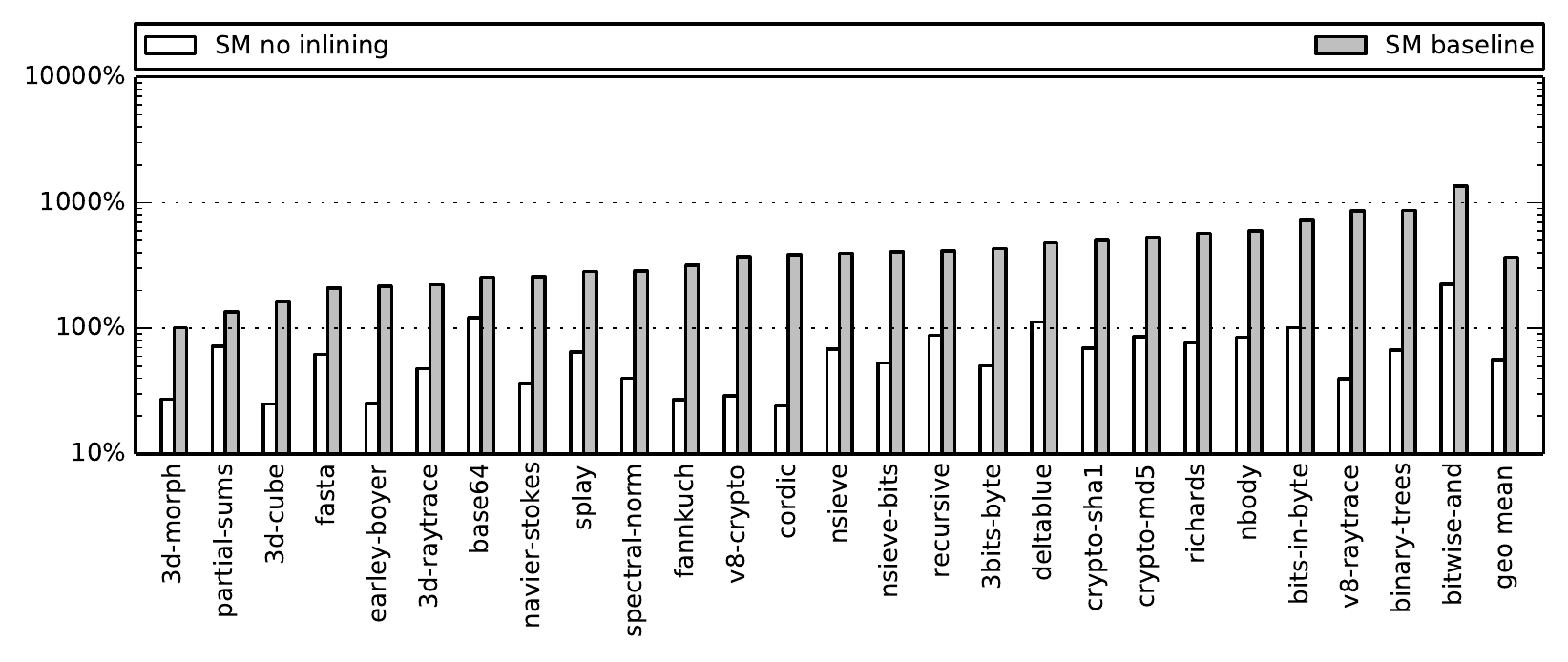}
    \end{center}
\caption{Speed relative to SpiderMonkey baseline and SpiderMonkey without inlining (log scale, bars above 100\% favor Higgs)\label{fig:against_sm}}
\end{figure*}

The execution time performance of Higgs was also compared to that of the
SpiderMonkey baseline compiler and the IonMonkey optimizing JIT (see
Figure~\ref{fig:against_sm}). The IonMonkey configuration tested has inlining,
loop unrolling and loop invariant code motion disabled to make its capabilities
more comparable to that of Higgs.

Higgs outperforms the SpiderMonkey baseline compiler by a wide margin, with
speedups of up to 1357\unskip\%. This confirms that
in terms of execution time, Higgs performs significantly better than a typical
baseline compiler. Higgs outperforms both Crankshaft and IonMonkey on
{\tt bitwise-and}, which again confirms that Higgs has faster global variable
access, thanks to typed shapes and shape propagation.

The performance of Higgs is close to that of IonMonkey on several benchmarks,
even outperforming it in a few cases. We believe that implementing a faster
calling convention for Higgs, as well as other improvements outlined in
Section~\ref{sec:future}, should make the performance of Higgs even more
competitive.

\subsection{Comparison with Truffle/JS}

\begin{figure*}[tb]
    \begin{center}
    \includegraphics[scale=1.00]{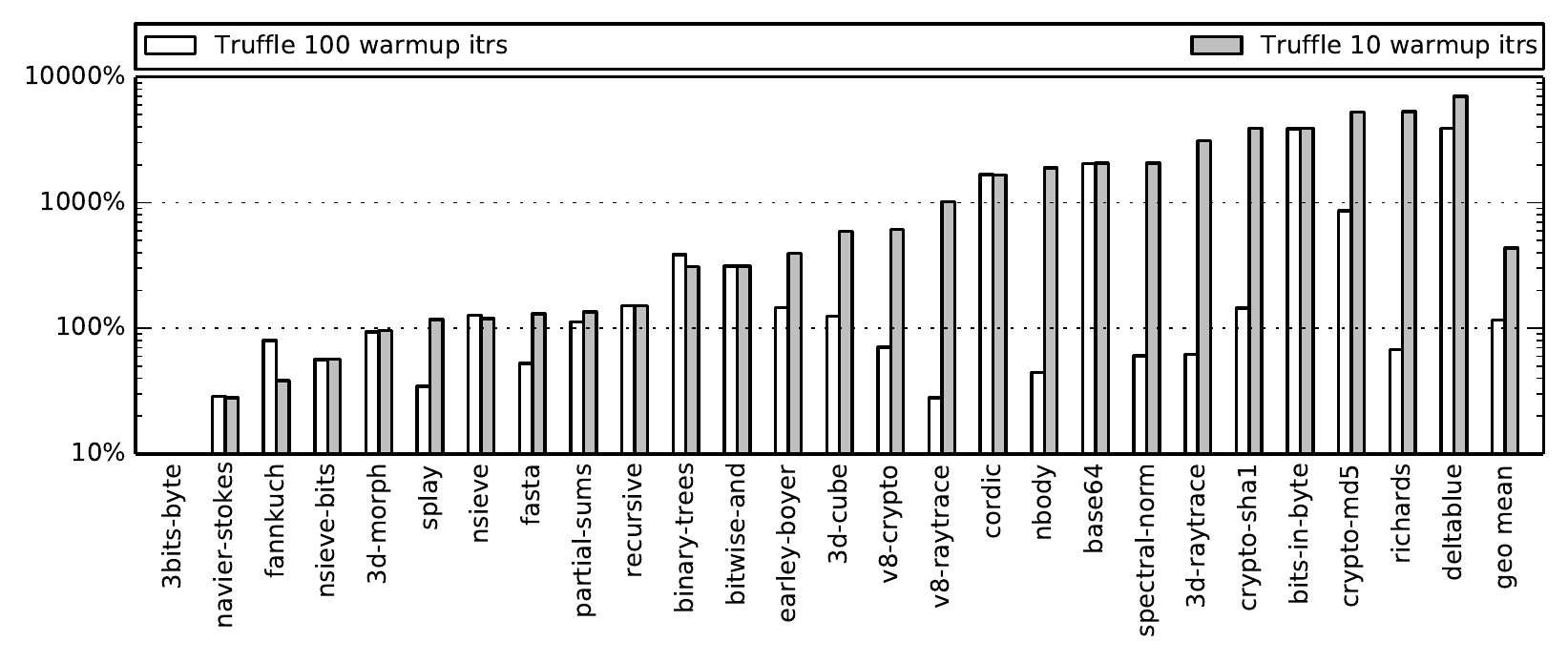}
    \end{center}
\caption{Speed relative to Truffle/JS (log scale, bars above 100\% favor Higgs)\label{fig:against_tr}}
\end{figure*}

Since Truffle has longer warmup times than other systems, we have compared
the performance of Higgs to that of Truffle with 10 and 100
warmup iterations. Higgs performs better than Truffle with 100 warmup iterations
on 13\unskip
out of 26\unskip benchmarks and yields an average speedup of
16\unskip\%.

Truffle has two main performance advantages over Higgs. The first is
that after warmup, Truffle is able to perform deep inlining, as illustrated by
the {\tt v8-raytrace} benchmark. The second is that Truffle has sophisticated
analyses which Higgs does not have. For instance, the recorded time for the
{\tt 3bits-byte} microbenchmark is zero, suggesting that Truffle was able to
entirely eliminate the computation performed as its output is never used. Doing
this requires a side-effect analysis which can cope with the semantic
complexities of JavaScript.

It is interesting to note that Higgs performs much better on the
{\tt bitwise-and} microbenchmark, indicating that Higgs has faster global
variable accesses than Truffle/JS, SpiderMonkey and V8. Higgs also outperforms
Truffle/JS on {\tt recursive}, which takes advantage of known callee
identities provided by typed shapes, as well as {\tt binary-trees}, which
makes heavy use of objects.

\section{Limitations and Future Work}
\subsection{Limitations\label{sec:limitations}}

We have shown that typed shapes and shape propagation yield significant
speedups on our benchmark set, and outlined a mechanism that effectively
prevents code size explosions. It is possible to imagine pathological cases
where an immense blowup in the number of object shapes might occur, and our
approach may lead to performance degradations. The results obtained, however,
suggest that such pathological cases are uncommon. Furthermore, it is not
difficult to imagine a mechanism to limit performance degradations on such
edge cases, by simply disabling the type-specialization of shapes for specific
shape subtrees. Hence, we stand by the proposition that typed shapes offer
attractive performance advantages.

At this point in time, Higgs lacks several crucial optimizations which would
be needed were it to become a viable commercial compiler and directly compete
against V8, SpiderMonkey and others. Notably, Higgs lacks function inlining, loop
unrolling, loop invariant code motion, and automatic vectorization. Higgs also
has relatively slow compile times since it is written in a garbage collected
programming language and has not been fine-tuned for fast compilation.

Higgs is a one-PhD-student effort, whereas V8 and TM have large teams
of expert software engineers to draw on. They can optimize their implementation
in ways we simply cannot. We believe that this paper offers a convincing
demonstration that typed shapes and shape propagation offer competitive
performance advantages in the realm of optimizing object property accesses.
The techniques outlined in this paper could benefit V8, SpiderMonkey and many
other JIT compilers for object-oriented dynamic languages.

\subsection{Future Work\label{sec:future}}

Work by Costa, Alves et al.~\cite{jit_value_spec} has shown that significant
speedups can be obtained by specializing JS code based on function
argument values, which are often constant. In a similar vein, typed shapes
could be extended to allow for the direct encoding of constant values into
object shapes. This would likely be particularly useful for global variables
which are never mutated and effectively constant.

An important limitation of the shape propagation approach as presented in this
paper is that it is intraprocedural only. Known shapes are now propagated to
callees, and furthermore, known shape information is lost whenever a
function call is made. This is because function calls are currently treated
as black boxes, and it is not guaranteed that callees will not change the shape
of objects used in the caller.

We believe it may be interesting to investigate interprocedural basic block
versioning. Namely, it may be useful to specialize function entry points so
that known types can be propagated from callers to callees. This would
contribute to eliminating more type tests and shape tests. Typed
shapes will make the implementation of interprocedural BBV easier and more
efficient, since they provide precise information about callee identities.

Having information about callee identities should also make it possible to
implement a rudimentary system to assess whether or not callees
modify object shapes or not. Having a way to guarantee that a callee will not
cause any object to change shape makes it possible to avoid discarding shape
information at call sites, thereby improving the effectiveness of shape
propagation.

Another inefficiency which may be useful to fix is the inefficient calling
convention used by Higgs. Arguments are currently passed through the stack.
Passing arguments directly in registers would make function calls
significantly faster, making the performance of Higgs more competitive.

\section{Related Work}


Polymorphic inline caches were originally introduced in literature discussing
the efficient implementation of the Self programming
language~\cite{self, self_pic}. Self did not use typed shapes exactly as
discussed in this paper, but instead a concept of {\em maps} which grouped
objects cloned from the same prototype. These served essentially the same
function as shapes, reducing memory usage overhead and storing metadata
relating to properties.


Commercial JS implementations such as Google's V8, Mozilla's
SpiderMonkey and Apple's SquirrelFish all make use of polymorphic inline
caches to speed up property accesses. These also make use of either object
shapes or something resembling maps. Oracle's {\em Truffle} framework for the
implementation of dynamic languages~\cite{trufflejs, truffle} uses a chaining
of dynamic tests equivalent to polymorphic inline caches to improve
performance.

The Truffle object storage model~\cite{truffle_obj} describes a typical
implementation of an object system where each object contains a pointer to
its shape, which describes the layout of the object (property locations) and
property attribute metadata. Property additions cause shape transitions.
Type tag information is stored in shapes and properties are unboxed.

An important difference with our approach is that Truffle only allows for
acyclic property type transitions, that is transition to wider types in the
type hierarchy. Typed shapes allow objects to switch between different shapes
so that property values are always unboxed.


Several {\em whole-program type analyses} for JS
were developed~\cite{tajs, tajs_lazy, type_ref}. These analyses are generally
considered too expensive to use in a JIT compiler. They also tend to suffer
from precision limitations when dealing with object types. It is often
difficult, for instance, to prove that a specific property of an object must be
initialized at a given program point. Not being able to prove this means that
every property access must assume the property could take the {\tt undefined}
value, which pollutes analysis results.

The work done by Kedlaya, Roesch et al.~\cite{impr_type_spec} shows strategies
for improving the precision of type analyses by combining them with type
feedback and profiling. This strategy shows promise, but does not explicitly
deal with object shapes and property types. Work has also been done on a
flow-sensitive alias analysis for dynamic languages~\cite{alias_dynamic}. This
analysis tries to strike a balance between precision and speed, it is likely
too expensive for use in JIT compilers, however.

Work done by Brian Hackett et al. at Mozilla resulted in an interprocedural
hybrid type analysis for JS suitable for use in production JIT
compilers~\cite{mozti}. This approach has a notion of object types segregating
objects by prototype, and tries to bound the possible types a given property
associated with a given object type may have. The Mozilla approach does not
always guarantee that a given property has a given type, and so often cannot
unbox property values. It is also limited when it comes to proving that
properties must exist, relying on a supplemental analysis which examines
constructor function bodies to try and prove initialization. The approach
we present is simpler and potentially more precise.


{\em Trace compilation}, originally introduced by the Dynamo~\cite{dynamo}
native code optimization system, and later applied to
JIT compilation in HotpathVM~\cite{hotpathvm} aims to record long sequences
of instructions executed inside hot loops. Such linear sequences of
instructions often make optimization simpler. Type information can be
accumulated along traces and used to specialize code and remove type
tests~\cite{trace_type_spec}, overflow checks~\cite{trace_ovf} or unnecessary
allocations~\cite{trace_alloc}.

The {\em TraceMonkey} tracing JIT compiler for JS can specialize traces
based on types~\cite{trace_type_spec}. It can also guard based on object
shapes and eliminate some shape dispatch overhead inside traces, similarly
to the shape propagation discussed in this paper. It does not, however
specialize code based on property types. Trace compilation~\cite{pypy} and
meta-tracing are an active area of research~\cite{bolz_tracing_racket} in the
realm of dynamic language optimization. Most tracing JIT compilers for
languages which have some concept of objects, tuples or records could likely
benefit from the approaches discussed in this paper.

Facebook's {\em HipHop VM} for PHP~\cite{hiphopvm} uses an approach called Tracelet
specialization which has many similarities with BBV. Seeing
that PHP is an object-oriented dynamic language and that HipHop VM already
specializes code using type guards, it seems this system could likely benefit
from typed shapes and shape propagation.


Grimmer, Matthias et al.~\cite{c_structs_js} implemented an interpreter which
can access C structs and arrays as JS objects at better speeds than native
JS objects. This is useful when interfacing with C, but likely impractical as
a drop-in replacement for JS objects.

The upcoming ECMAScript 6 specification~\cite{es6_spec} will include {\em typed
arrays}, which are arrays constrained to contain uniformly typed elements
(for example, 8-bit signed integer arrays).

There is a proposal for the inclusion of {\em typed objects} (also known as
"struct types") in ECMAScript 7, a future revision of the JS
specification. These are objects using pre-declared memory layouts with with
type-annotated fields, much in the way one would declare a {\t struct} in C.
One of the stated goals is to improve optimization opportunities for JIT
compilers~\cite{typed_objects}. The work presented in this paper aims to bring
much of the performance advantages of typed objects without requiring the
programmer to declare explicit type annotations or fixed object layouts.

\section{Conclusion}\label{sec:conclusion}
We have described two techniques to effectively specialize code based on object
and property types. Typed shapes, an extension to the familiar object shapes
used in most commercial JavaScript engines, enables us to extract property
type information on property reads. Shape propagation allows us to propagate
object shapes as code is generated, reducing the overhead of Polymorphic
Inline Caches (PICs).

Across the 26\unskip benchmarks tested, these techniques eliminate on
average 48\unskip\% of type tests and
45\unskip\% of shape tests. Code size is reduced
by 17\unskip\% and execution time by
25\unskip\%. Our results also show that Higgs
has faster global variable accesses than Truffle/JS, SpiderMonkey and V8,
likely because Higgs uses typed shapes to manipulate the global object as any
other object, and this provides excellent performance.

The techniques presented are simple to implement and combine particularly well
with a compiler architecture based on Basic Block Versioning (BBV), but
should be easily adaptable to compilers based on trace compilation or
method JITs with type feedback. An unseen benefit of shape propagation is that
it provides aliasing information, which may be very useful for many kinds of
optimizations beyond the scope of this publication.

Higgs is open source and the code used in preparing this publication is
available on GitHub\footnote{https://github.com/higgsjs/Higgs/tree/dls2015}.

\acks

Special thanks go to Laurie Hendren, Erick Lavoie, Vincent Foley, Paul Khuong,
Tommy Everett, Brett Fraley and all those who have contributed to the
development of Higgs.

This work was supported, in part, by the Natural Sciences and Engineering
Research Council of Canada (NSERC) and Mozilla Corporation.

\bibliographystyle{plain}
\bibliography{main}

\end{document}